\documentclass[12pt]{iopart}

\begin{document}

\title{The Chiral MagnetoHydroDynamics of QCD fluid at RHIC and LHC}

\author{Dmitri E. Kharzeev$^{1,2}$}

\address{$^1$ Department of Physics and Astronomy,\\
Stony Brook University, Stony Brook, New York 11794-3800, USA\\
$^2$ Physics Department, Brookhaven National Laboratory,\\
Upton, New York 11973-5000, USA}
\ead{Dmitri.Kharzeev@stonybrook.edu}

\begin{abstract}
The experimental results on heavy ion collisions at RHIC and LHC indicate that QCD plasma behaves as a nearly perfect fluid described by relativistic hydrodynamics. 
Hydrodynamics is an effective low-energy Theory Of Everything stating that the response of a system to external perturbations is dictated by conservation laws that are a consequence of the symmetries of the underlying theory. 
In the case of QCD fluid produced in heavy ion collisions, this theory possesses anomalies, so some of the apparent classical symmetries are broken by quantum effects. 
Even though the anomalies appear as a result of UV regularization and so look like a 
short distance phenomenon, it has been realized recently that they also affect the large distance, macroscopic behavior in hydrodynamics. One of the manifestations of anomalies in relativistic hydrodynamics is the Chiral Magnetic Effect (CME). At this conference, a number of evidences for CME have been presented, including i) the disappearance of charge asymmetry fluctuations in the low-energy RHIC data where the energy density is thought to be below the critical one for deconfinement; ii) the observation of charge asymmetry fluctuations in Pb-Pb collisions at the LHC. Here I give a three-page summary of some of the recent theoretical and experimental developments and of the future tests that may allow to establish (or to refute) the CME as the origin of the observed charge asymmetry fluctuations.
\end{abstract}

\maketitle

\section{Introduction}
Recently it has become clear that quantum anomalies play a very important role in the macroscopic dynamics of relativistic fluids. Much of this progress is motivated by the possibility to observe the anomalous "chiral magnetic" currents in non-Abelian quark-gluon plasma produced at RHIC and LHC. 

The Chiral Magnetic Effect (CME) is the anomaly--induced phenomenon of  electric charge separation along the axis of the applied magnetic field in the presence of fluctuating topological charge \cite{Kharzeev:2004ey,Kharzeev:2007tn,Kharzeev:2007jp,Fukushima:2008xe,Kharzeev:2009fn}. The CME in QCD coupled to electromagnetism assumes a chirality asymmetry
between left- and right-handed quarks, parameterized by a chiral 
chemical potential $\mu_A$.  Such an asymmetry can arise if there is
an asymmetry between the 
topology-changing transitions early in a heavy ion
collision.  
Closely related phenomena have been discussed in the physics of neutrinos  \cite{Vilenkin:1979ui}, 
primordial electroweak plasma \cite{Giovannini:1997gp} and quantum
wires \cite{acf}.  

\section{Recent theoretical developments}

\subsection{The absence of CME current renormalization at weak coupling} 

The longitudinal part of the axial anomaly does not receive perturbative corrections, 
so it was anticipated that the formula for CME current would not be subject to perturbative renormalization. This has been established in the careful field theoretical analysis in Refs \cite{Hong:2010hi,Hou:2011ze}. 

\subsection{The absence of CME current renormalization at strong coupling: the holography}

While the original derivations used the weak coupling methods, the origin of the effect is essentially topological and so the CME is not renormalized even at (infinitely) strong coupling, as was shown by the holographic methods  \cite{Yee:2009vw,Rubakov:2010qi,Rebhan:2009vc,Gynther:2010ed,Gorsky:2010xu,Brits:2010pw}. For a sample of recent interesting developments in holographic approaches to CME and related phenomena, see Refs. \cite{Kalaydzhyan:2011vx, Landsteiner:2011cp,Hoyos:2011us,Mateos:2011tv,Landsteiner:2011iq} and references therein.

\subsection{Lattice QCD $\times$ QED}

The evidence for the CME has been found in lattice QCD coupled to electromagnetism, both within the quenched approximation \cite{Buividovich:2009wi,Buividovich:2009zzb,Buividovich:2010tn} and with light domain wall fermions \cite{Abramczyk:2009gb}. Unlike the baryon chemical potential, the chiral chemical potential $\mu_A$ does not present a ``sign problem"  which opens a possibility for lattice computations at finite $\mu_A$ \cite{Fukushima:2008xe}. A direct lattice study of the chiral magnetic current as a function of $\mu_A$ was performed very recently \cite{Yamamoto:2011gk}; it confirms the expected dependence of CME on the chiral chemical potential and the magnetic field. 

\subsection{Chiral MagnetoHydroDynamics}

Since in the strong coupling regime the plasma represents a fluid, it is of great interest to study the effects 
of anomalies in relativistic hydrodynamics. 
A purely hydrodynamical derivation of the anomaly effects at first order in the derivative expansion was given by Son and Surowka \cite{Son:2009tf}. Their idea is to impose on the local entropy production rate $\partial_\mu s^\mu$ the positivity constraint reflecting the second law of thermodynamics. 
This leads to a set of algebro-differential equations for the transport coefficients related to the anomaly.
It has been found that the CME current persists in hydrodynamics \cite{Son:2009tf, Sadofyev:2010pr,Sadofyev:2010is,Kalaydzhyan:2011vx} and is transferred by the sound-like gapless excitation -- ``the chiral magnetic wave" \cite{Kharzeev:2010gd,Burnier:2011bf}, see also \cite{Newman:2005hd}.


The full second order formulation of anomalous effects in relativistic hydrodynamics with external electromagnetic fields (``Chiral MagnetoHydroDynamics", CMHD) was recently given in Ref. \cite{Kharzeev:2011ds}. The guiding principle proposed in Ref. \cite{Kharzeev:2011ds} is the following: the ${\cal T}$-even anomaly-induced viscous terms must not contribute to the entropy production, as ${\cal T}$-even conductivities describe {\it non-dissipative} currents. 
This allows to compute analytically 13 out of 18 anomaly-related transport coefficients in second order conformal hydrodynamics. Explicit holographic computations confirm the validity of these results.

\section{Recent experimental developments}

Recently, STAR \cite{:2009uh} and PHENIX \cite{phenix}
Collaborations at Relativistic Heavy Ion Collider at BNL reported
experimental observation of charge asymmetry fluctuations.  
At this conference, STAR reported \cite{Mohanty:2011nm} the expected 
\cite{Kharzeev:2007jp,Fukushima:2008xe} disappearance of the effect at low collision energies where the energy density of created matter is smaller and likely below the critical one needed for the restoration of chiral symmetry. The first analysis at the Large Hadron Collider has been performed by the ALICE Collaboration that has reported \cite{Collaboration:2011sm} the observation of charge asymmetry fluctuation signaling the persistence of the effect at very high energy densities. The magnitude of the effect is similar to the one observed at RHIC, likely reflecting the same magnetic flux through QCD fluid and close ratio of the topological charge and energy densities \cite{Kharzeev:2007jp,Fukushima:2008xe}.

\section{Future tests}

Additional future tests of CME include the positive correlation between the electric and baryon charge asymmetries \cite{Kharzeev:2010gr}, different in strength for chiral magnetic and chiral vortical \cite{Kharzeev:2007tn,Son:2009tf,Rogachevsky:2010ys} effects; see also \cite{Voloshin:2010ut,KerenZur:2010zw} for other ideas.

\vskip 0.3cm 
I am indebted to my collaborators for sharing their insights with me. Space limitations do not allow me to describe all important developments in this rapidly evolving field.  
 This work was supported by the U.S. Department of Energy under Contracts No.
DE-AC02-98CH10886 and DE-FG-88ER41723.

\end{document}